\documentclass[twocolumn,showpacs,preprintnumbers,prl,amsmath,amssymb,superscriptaddress,graphics]{revtex4-1}
\usepackage{graphicx}
\usepackage{amsmath}
\usepackage{amssymb}
\usepackage[colorlinks,linkcolor=blue,
            citecolor=blue,
            hyperindex,
            pdfstartview=FitH,
            plainpages=false]
            {hyperref}
\usepackage{color}

\newcommand{\beq}{\begin{equation}}
\newcommand{\eneq}{\end{equation}}

\begin{document}

\title{Crystalline symmetry-protected non-trivial topology in prototype compound BaAl$_4$}

\author{Kefeng Wang\textsuperscript{+}\footnotetext[0]{\textsuperscript{+} These authors contributed equally to this work.}}
\affiliation{Center for Nanophysics and Advanced Materials, Department of Physics, University of Maryland, College Park, MD 20742, USA}
\author{Ryo Mori\textsuperscript{+}}
\affiliation{Applied Science \& Technology, University of California, Berkeley, CA 94720, USA}
\affiliation{Materials Sciences Division, Lawrence Berkeley National Laboratory, Berkeley, CA 94720, USA}
\author{Zhijun Wang}
\affiliation{Beijing National Laboratory for Condensed Matter Physics, and Institute of Physics, Chinese Academy of Sciences, Beijing 100190, China}
\affiliation{University of Chinese Academy of Sciences, Beijing 100049, China}
\author{Limin Wang}
\affiliation{Center for Nanophysics and Advanced Materials, Department of Physics, University of Maryland, College Park, MD 20742, USA}
\author{Jonathan Han Son Ma}
\affiliation{Materials Sciences Division, Lawrence Berkeley National Laboratory, Berkeley, CA 94720, USA}
\affiliation{Department of Physics, University of California, Berkeley, CA 94720, USA}
\author{Drew W. Latzke}
\affiliation{Applied Science \& Technology, University of California, Berkeley, CA 94720, USA}
\affiliation{Materials Sciences Division, Lawrence Berkeley National Laboratory, Berkeley, CA 94720, USA}
\author{David E. Graf}
\affiliation{National High Magnetic Field Laboratory, Florida State University, Tallahassee, FL 32306-4005, USA}
\author{Jonathan D. Denlinger}
\affiliation{Advanced Light Source, Lawrence Berkeley National Laboratory, Berkeley, CA 94720, USA}
\author{Daniel Campbell}
\affiliation{Center for Nanophysics and Advanced Materials, Department of Physics, University of Maryland, College Park, MD 20742, USA}
\author{B. Andrei Bernevig}
\affiliation{Department of Physics, Princeton University, Princeton, NJ 08554, USA}
\author{Alessandra Lanzara}
\email[To whom correspondence should be addressed. Email: ]{lanzara@berkeley.edu}
\affiliation{Materials Sciences Division, Lawrence Berkeley National Laboratory, Berkeley, CA 94720, USA}
\affiliation{Department of Physics, University of California, Berkeley, CA 94720, USA}
\author{Johnpierre Paglione}
\email[To whom correspondence should be addressed. Email: ]{paglione@umd.edu}
\affiliation{Center for Nanophysics and Advanced Materials, Department of Physics, University of Maryland, College Park, MD 20742, USA}
\date{\today}

\begin{abstract}
The BaAl$_4$ prototype crystal structure is the most populous of all structure types, and is the building block for a diverse set of sub-structures including the famous ThCr$_2$Si$_2$ family that hosts high-temperature superconductivity and numerous magnetic and strongly correlated electron systems. The MA$_4$ family of materials (M=Sr, Ba, Eu; A=Al, Ga, In) themselves present an intriguing set of ground states including charge and spin orders, but have largely been considered as uninteresting metals. Using electronic structure calculations, symmetry analysis and topological quantum chemistry techniques, we predict the exemplary compound BaAl$_4$ to harbor a three-dimensional Dirac spectrum with non-trivial topology and possible nodal lines crossing the Brillouin zone, wherein one pair of semi-Dirac points with linear dispersion along the $k_z$ direction and quadratic dispersion along the $k_x/k_y$ direction resides on the rotational axis with $C_{4v}$ point group symmetry. Electrical transport measurements reveal the presence of an extremely large, unsaturating positive magnetoresistance in BaAl$_4$ despite an uncompensated band structure, and quantum oscillations and angle-resolved photoemission spectroscopy measurements confirm the predicted multiband semimetal structure with pockets of Dirac holes and a Van Hove singularity (VHS) remarkably consistent with the theoretical prediction. We thus present BaAl$_4$ as a new topological semimetal, casting its prototype status into a new role as building block for a vast array of new topological materials.
\end{abstract}

\maketitle


The development of a better understanding of the role of topology in electronic states has led to a revolution in exploring both old and new compounds for non-trivial topologies, and has provided a fertile platform to search for new topological fermionic excitations~\cite{TIReview1,TIReview2,Armitage2018}. 
The discovery of topological insulators (TIs) with robust metallic surface states protected by the topology of the insulating bulk~\cite{TIReview1,TIReview2} was followed by the discovery of topological semimetals (TSMs). These are found to have deep connections with particle physics models of relativistic chiral (Weyl) fermions and also provide fermionic excitations without high-energy counterparts~\cite{Young_2012,Wan_2011,Fang_2012,NodalChain,NodalLine,Triple-1,Triple-2,MoP-ARPES,EightFold,Armitage2018}. In a Dirac semimetal (DSM), bulk valence and conduction bands cross each other linearly at discrete points protected by time-reversal, inversion and crystal symmetries. These stable Dirac points with four-fold degeneracy make it possible to simulate massless Dirac fermions~\cite{Young_2012,Armitage2018}. By breaking either inversion or time reversal symmetry, a Dirac semimetal can be tuned to a Weyl state where the nondegenerate linear touchings of the bulk bands come in pairs. Each Weyl point with two-fold-degeneracy has a definite chirality of $\pm1$ and the quasiparticle excitations mimic the relativistic Weyl (chiral) fermions~\cite{Wan_2011,Fang_2012,TaAs-ARPES1,TaAs-ARPES2}.

Cd$_3$As$_2$ and Na$_3$Bi were the first theoretically predicted DSM candidates~\cite{A3Bi,Cd3As2-1}, subsequently confirmed by experiments~\cite{Borisenko_2014,Neupane_2014,Liu2014,Jeon_2014}.
Recently, several materials including transition-metal icosagenides MA$_3$ (M=V, Nb, and Ta; A=Al, Ga, In)~\cite{VAl3}, MgTa$_2$N$_3$~\cite{MgTa2N3,MgTa2N3-1}, CaAgBi~\cite{CaAgBi} and PtTe$_2$~\cite{PtTe2}, were predicted to host Dirac fermions. These form part of an even larger list of TSMs \cite{Zhang2018,TMDatabase,TMDatabasewebsite} that also exhibit exotic physical properties including high carrier mobilities, chiral anomaly physics and topological superconductivity. Extremely large magnetoresistance (MR) is another symptom of such materials, which is usually very small in non-magnetic, non-compensated metals according to semiclassical transport theory. However, several TSMs -- including Dirac semimetals Cd$_3$As$_2$ and ZrSiS~\cite{Liang_2014,Narayanan_2015,ZrSiS-MR}, the NbAs family~\cite{Shekhar_2015,TaAs-Jia,TaAs-Chen,NbAs-Luo}, the type-II Weyl semimetal WTe$_2$~\cite{WTe2-MR} and several other TSMs such as NbSb$_2$~\cite{NbSb2-Wang}, PtSn$_4$~\cite{PtSn4-MR,PtSn4-ARPES}, LaSb\cite{LaSb_Tafti} -- exhibit extremely large MR behavior which could have potential application in magnetic sensors, spintronics and memories. Very recently, multiple topological states have been predicted and observed in the iron-based superconductors LiFeAs and Fe(Se,Te)\cite{Iron_topological}, while possible Dirac fermions are predicted in heavy fermion superconductors Ce(Co,Rh,Ir)In$_5$\cite{CeCoIn5_topology}.

\begin{figure*}
	\begin{center}
		\includegraphics[clip,width=\linewidth]{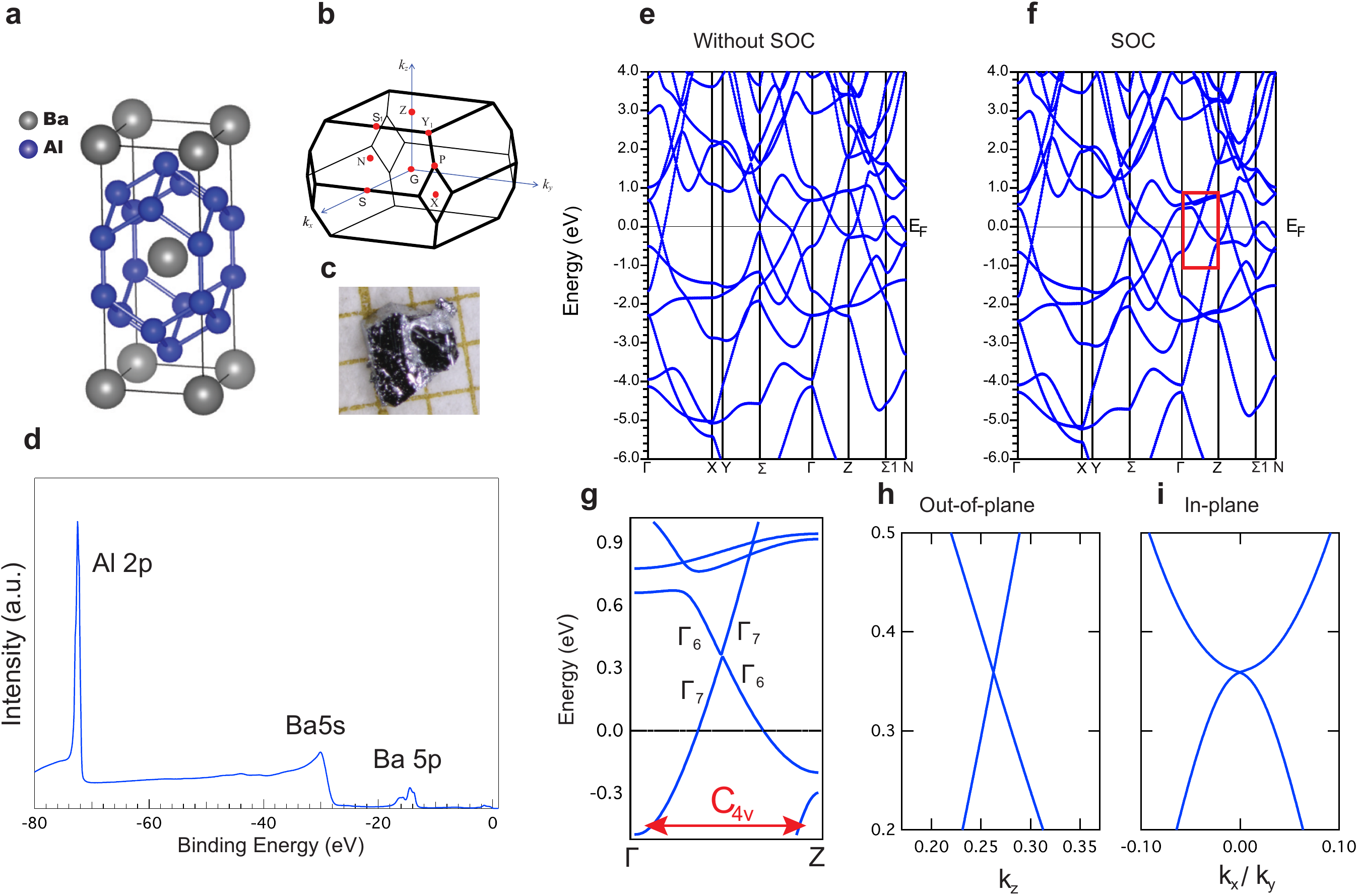}
		\caption{
			\textbf{Non-trivial topology of prototype compound BaAl$_4$.}
			The crystal structure (a) and Brillouin zone (b) of single-crystalline BaAl$_4$ (c) with $I4/mmm$ space group are presented along with 		
			electronic structure calculations and symmetry analysis, revealing a three-dimensional Dirac spectrum with non-trivial topology.
			(d) The integrated photoemission core-level spectrum exhibits sharp peaks from Ba-5$s$, Ba-5$p$, and Al-2$p$ core-levels, confirming high 
			sample quality.
			The calculated bulk electronic structure is presented (e) without spin-orbit coupling and (f) with spin-orbit coupling included, with red box highlighting the $\Gamma$--$Z$ dispersion where a Dirac point is located. 
			(g) Enlarged low-energy band structure with irreducible representation, where the two crossing bands belong to two different irreducible representations of the $C_{4v}$ point group, showing the accidental band crossing protected by the crystal symmetry.
			The zoomed 3D Dirac structure exhibits a linear dispersion along the out-of plane ($k_z$) direction [panel (h)] and a quadratic dispersion  along the in-plane ($k_x$/$k_y$) directions [panel (i)], classifying this as a semi-Dirac point where massless and massive fermions can coexist at the same point in momentum space (see text).}
	\end{center}
\end{figure*}

\begin{figure*}
	\begin{center}
		\includegraphics[clip,width=\linewidth]{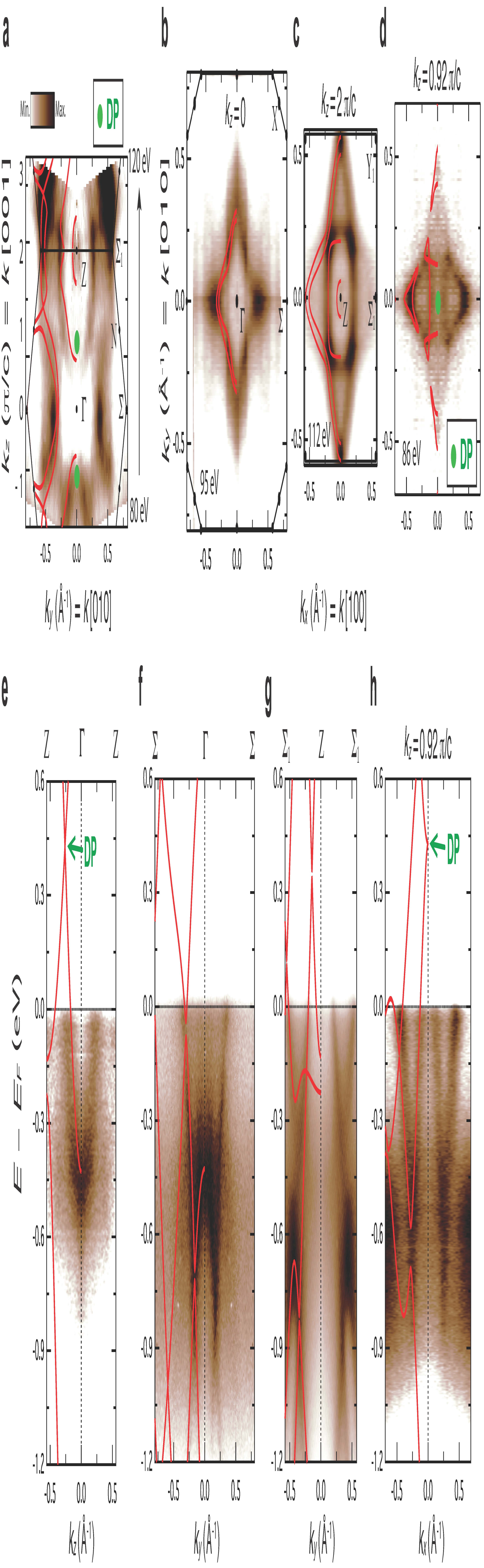}
		\caption{
			\textbf{ARPES spectra near the Fermi level compared to DFT calculation of BaAl$_4$.}
			\textbf{a--c,} Constant energy maps at {\sl E$_F$} in the high-symmetry planes: the $k_y-k_z$ plane (perpendicular to the surface (001)-plane) (\textbf{a}) using the incident photon energy ranging from 80 to 120 eV, the $k_x-k_y$ plane crossing $\Gamma$ (\textbf{b}) using the incident photon energy of 95 eV, and the $k_x-k_y$ plane crossing $Z$ (\textbf{c}) using the incident photon energy of 112 eV. Black solid line represents the bulk Brillouin zones (BZ), marked with high-symmetry points.
			\textbf{d,} Constant energy map at {\sl E$_F$} crossing the location of Dirac point (DP) using the incident photon energy of 86 eV.
			\textbf{e--g,} ARPES spectra of energy versus momentum cuts along the high-symmetry directions: $Z-\Gamma-Z$ (\textbf{e}), $\Sigma-\Gamma-\Sigma$ (\textbf{f}), and $\Sigma_1-Z-\Sigma_1$ (\textbf{g}).
			\textbf{h,} ARPES spectra of energy versus momentum cut crossing the DP along the in-plane direction $k_y$.
			All spectra shown in \textbf{a--d} are obtained by integrating intensities in the energy window of {$\pm20$} meV at {\sl E$_F$}.
			The red lines superposed in all panels here are the DFT calculation results for BaAl$_4$.
			The locations of Dirac point are marked with green dots in \textbf{a,d}.
			For a direct comparison, all Fermi surfaces (\textbf{a--d}) and all energy and momentum cuts (\textbf{e--h}) shown here are centered at $(k_x, k_y, k_z)=(0, 0, 0)$.
		}
	\end{center}
\end{figure*}

The MA$_4$ family of materials (M=Sr, Ba, Eu; A=Al,Ga,In) form the prototype parent crystal structure of a large family of compounds including the ThCr$_2$Si$_2$-based iron superconductors\cite{Paglione_iron_review}, classic heavy fermions systems such as CeCu$_2$Si$_2$\cite{CeCu2Si2,HeavyFermion_Review}, and numerous non-centrosymmetric compounds\cite{BaAl4_Derived}. In addition, the MA$_4$ family itself presents a rich variety of ground states. For instance, SrAl$_4$ exhibits a charge density wave transition at $200$ K, while Eu(Ga,Al)$_4$ orders antiferromagnetically at 15 K and is thought to also exhibit charge density wave order~\cite{SrAl4_FS,EuGa4}. 
In this work, we present the first comprehensive combined experimental and theoretical study of topology in the classic prototype compound BaAl$_4$, including first principles electronic structure calculations, transport measurements, and angle-resolved photoemission (ARPES) studies. We predict BaAl$_4$ to be a topological three dimensional Dirac semimetal, where one pair of type-I Dirac points resides on the rotational axis with the $C_{4v}$ point group symmetry, and confirm this prediction using ARPES and quantum oscillations experiments. The calculated band structure is in excellent agreement with ARPES spectra, and Hall resistivity and quantum oscillations measurements confirm the multi-band structure with electron and hole pockets, and is consistent with the predicted presence of Dirac hole pockets. Together with an extremely large MR that does not saturate in fields up to 32 T, our results indicate BaAl$_4$ is a new topological semimetal, and point to the MA$_4$ family of materials as fertile territory for the exploration of a new basis of topological systems.

\subsection{Non-Trivial Topology in BaAl$_4$}

Fig. 1(a,b) presents schematic illustrations of both the crystal structure and the bulk Brillouin zone of BaAl$_4$. The high quality of single crystals grown by a self-flux method (see Fig. 1(c)) is confirmed by X-ray diffraction (FIG. S1 in SM) and core-level electron photoemission spectrum, as shown in Fig. 1(d). Rietveld refinement of powder X-ray diffraction spectra confirms our samples to have the body-centered tetragonal structure with space group $I4/mmm$ (No. 139). Clear peaks from Ba-5$s$, Ba-5$s$, and Al-2$p$ core-level electrons observed by photoemission spectroscopy confirm the correct chemical composition. 

\begin{figure*}
	\begin{center}
		\includegraphics[clip,width=\linewidth]{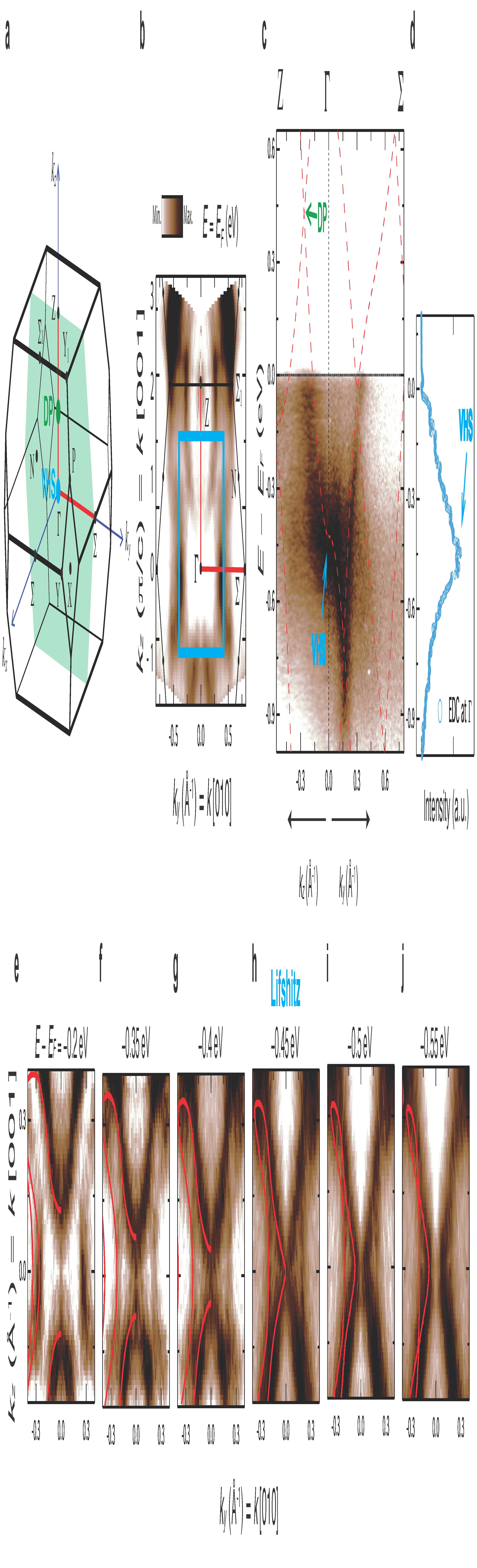}
		\caption{
			\textbf{Lifshitz point and Van Hove singularity in BaAl$_4$.}
			\textbf{a,} The bulk BZ marked with the location of Dirac point (DP, green dot) and Van Hove singularity of saddle point type (VHS, light blue point). The light green plane represents the location for the constant energy contours shown in \textbf{b,e--j}. 
			\textbf{b,} Constant energy map at {\sl E$_F$}. The light blue square shows the location for \textbf{e--j}. Red lines in \textbf{a,b} are the high symmetry lines for the energy and momentum cut shown in \textbf{c}.
			\textbf{c,} ARPES spectra of energy versus momentum cuts along the $Z-\Gamma-\Sigma$. The local maximum and minimum are observed at the same momentum $\Gamma$ point, showing the saddle point singularity VHS.
			\textbf{d,} Energy-distribution curve (EDC) obtained by integrating the ARPES intensity in the vicinity of $\Gamma$ point (window size: $\pm$ 0.02 \AA$^{-1}$). The peak of the EDC marks the VHS at $E_B\sim-450$meV where the topology of Fermi surface changes.
			\textbf{e--j,} Series of the symmetrized constant energy maps at $E_B=-0.2$, $-0.35$, $-0.4$, $-0.45$, $-0.5$, and $-0.55$ eV, respectively, showing the topology of electronic structure changes at the Lifshitz point (\textbf{h}), $E_B\sim-450$meV.
			The red lines superposed in these panels here are the DFT calculation results for $E_B=-0.27$, $-0.4$, $-0.45$, $-0.497$, $-0.55$, and $-0.57$, respectively.
			For a direct comparison, all constant energy maps (\textbf{a,e--j}) and the energy and momentum cut (\textbf{c}) shown here are centered at $(k_x, k_y, k_z)=(0, 0, 0)$.
		}
	\end{center}
\end{figure*}

In the BaAl$_4$ unit cell, the Ba atoms sit at the Wyckoff position $2a$, while the Al atoms sit at two non-equivalent Wyckoff positions $4d$ and $4e$, as shown in Table I in the Supplementary Materials. Among them, $2a$ and $4d$ sites are the maximal Wyckoff positions, but $4e$ is not. The 4e Wyckoff position has a stabilizer group~\cite{TopologicalQuantumChemisty} which is a subgroup of the stabilizer group of the maximal positon $2a$ or $2b$. Fig. 1(e,f) show the calculated bulk electronic structure along high-symmetry lines without (with) spin-orbit coupling, showing its semimetallic ground state feature. One observes that there are several crossing points along $\Gamma-X$, $\Gamma-\Sigma$, and $Z-\Sigma_1$ directions in different mirror reflection planes from two quasi-linear bands close to the Fermi level ($E_F$) in the band structure without SOC. These Dirac nodal points could give rise to a variety of nodal Dirac nodal lines in each mirror plane and also exotic surface states, which deserve further study.  

Once the SOC is included, most of these crossing points open small gaps because the two crossing bands belong to same irreducible representation of the point group indicated by the eigenvalue analyses. For example, the two bands crossing each other along $\Gamma-\Sigma$ direction at $\sim0.1$ eV both belong to $\Gamma_5$ representation and then the crossing point is gapped out. However, the crossing point along $\Gamma-Z$ direction is different. On $\Gamma-Z$ path which is along the rotational axis, the point group symmetry is of $C_{4v}$. The two doubly-degenerate crossing bands (highlighted by the red square in Fig. 1(f)) are found to form two distinct irreducible representations $\Gamma_6$ and $\Gamma_7$ representation of the $C_{4v}$ double group. In Fig. 1(g), we show these two bands belong to different representations of the symmetry group; therefore, the intersection of the two double-degenerate bands is protected by the crystalline symmetry, $C_{4v}$, forming a massless Dirac fermion. Figs. 1(h,i) show the zoomed dispersion along out-of-plane ($k_z$) and in-plane directions ($k_x$/$k_y$). While the slopes of the two crossing bands along the out-of-plane ($k_z$) have opposite signs, reminiscent of a type-I Dirac point~\cite{Cd3As2-1,Borisenko_2014,Neupane_2014,Liu2014,Jeon_2014}, the dispersion along the other directions show quadratic dispersion. The crossing points are above the Fermi level, according to the ab initio calculations. The physical elemental band representations (pEBRs) are constructed by the orbitals at the maximal positions with time reversal symmetry~\cite{TopologicalQuantumChemisty}, with all the pEBRs of $2a$ $2b$ and $4d$ listed in Table II in the Supplementary Materials. For high-symmetry points, the eigenvalues (irreps) of all symmetry operators have been computed~\cite{Vergniory2018} from the VASP wave functions for each band~\cite{wzjcode,TMDatabase}. Referring to the character tables in the Bilbao Crystallographic Server (BCS)~\cite{TopologicalQuantumChemisty,server}, we successfully assign all the bands to corresponding irreducible representations (irreps). The irrep characters of the valence bands are listed in Table III in the Supplementary Materials. One can check if those valence bands can be decomposed into a linear combination of pEBRs. It turns out they can not, which means those valence bands cannot be topologically trivial.\\

\subsection{Angle-resolved photoemission}

To verify band structure calculations, we measured ARPES to obtain the electronic structure of BaAl$_4$ at various photon energies, using in-situ cleaved single crystals of BaAl$_4$ where the (001) surface cleavage plane is determined by X-ray diffraction (FIG. S1(b) of Supplementary Information).
Fig. 2 shows the measured ARPES spectra compared to calculation (red curves). The intensity variation across the measured ARPES data is attributed to photoemission matrix element effects. Note that we shifted the calculated band structure by 70 meV (hole doping) to take into account the experimental data in Fig.~2. In Fig.~2(a), we present the Fermi surface in the out-of-plane direction. The highly dispersive Fermi surface is clearly resolved, suggesting its bulk three-dimensional nature. Fig. 2(e) shows the energy and momentum cut along the $Z-\Gamma-Z$ line where the $C_{4v}$ exists. Consistent with the DFT prediction, the tail of the Dirac dispersion, which consists of small and large electron pockets at $Z$ and $\Gamma$ points respectively, is observed. To further investigate the electronic structures, we show the in-plane Fermi surfaces for the high-symmetry planes ($k_z\sim0$, $k_z\sim2\pi/c$, and $k_z\sim0.92\pi/c$ where the Dirac point (DP) protected by $C_{4v}$ is located) and the electronic dispersions ($\Gamma-\Sigma$, $Z-\Sigma_1$, and DP) in Fig. 2(b--c) and Fig. 2(f--h). Overall, the experimental Fermi surfaces and electronic dispersions are consistent with band structure calculations for all directions.

Similar to graphene, multiple Dirac dispersions in BaAl$_4$ create a saddle-point singularity in between the two DPs in momentum space as a result of a change in electronic topology, reminiscent of a Lifshitz transition in the electronic structure. Fig. 3(a) shows the location of the DP and VHS in the bulk BZ of BaAl$_4$. To identify the saddle-point singularity experimentally, we show the energy and momentum dispersion along $Z-\Gamma-\Sigma$ direction (along the red line in the Fig. 3(a,b)) in Fig. 3(c). While the tail of the Dirac dispersion has the local energy minimum in between the DPs, the $\Gamma$ point, the dispersion along the orthogonal direction (in-plane direction, $\Gamma-\Sigma$) has the opposite curvature, resulting in the saddle point singularity of band structure, a Van Hove singularity (VHS). The integrated energy distribution curve (EDC) around the $\Gamma$ point shown in Fig. 3(d) clearly indicates the peak of local density of states (DOS) at $\sim-$0.45 eV. This peak is consistent with a local DOS maximum in a saddle-point since the band structure is nearly flat at the VHS. Emergence of a VHS is associated with a Lifshitz transition. Fig. 3(e--j) represent a series of ARPES constant energy maps near the $\Gamma$ point (see the blue square in Fig. 3(b)) with various binding energies, emphasizing the change of topology of electronic structures. The energy contours of Dirac dispersions become close to each other toward the $\Gamma$ point as the binding energy changes to deeper level (shown in Fig. 3(e--g)). Once it reaches the Lifshitz point ($\sim-$0.45 eV), the two unconnected tails of Dirac dispersions merge and the electronic topology changes as shown in Fig. 3(h--j). This change in electronic topology is also clearly reproduced by the theoretical calculation (The red lines in Fig. 3(h--j)).
 
Because the Dirac point protected by $C_{4v}$ symmetry is above $E_F$ and hence not accessible to ARPES, we cannot directly observe the vicinity of DP. However, the excellent agreement between our ARPES results and the first-principles calculation, validates the theoretical prediction of Dirac fermions above $E_F$, confirming BaAl$_4$ as a topological semimetal with Dirac points. The observation of the saddle point singularity VHS and the Lifshitz point in between DPs identifies this family of materials as an ideal 3D analogue of graphene, similar to Na$_3$Bi. Moreover, in BaAl$_4$, while the Dirac dispersion along the out-of-plane axis ($k_z$) has a linearly type-I dispersion, the dispersion is quadratic along the in-plane axis ($k_x$/$k_y$) as shown in Fig. 1(h,i) and Fig. 2(e,h). Hence these DPs can be classified as semi-Dirac points where massless and massive fermions can coexist at the same point in momentum space. Similar semi-Dirac points are theoretically predicted and/or experimentally observed in two-dimensional systems such as a silicene oxide Si$_2$O and black phosphorus~\cite{blackphosphorus,Si2O}, and could result in interesting properties including highly anisotropic transport, distinct Landau level spectrum in a magnetic field, non-Fermi liquid, and Bloch-Zener oscillations.\cite{SemiDirac1,SemiDirac2}

\begin{figure*}
	\begin{center}
		\includegraphics[scale=0.6]{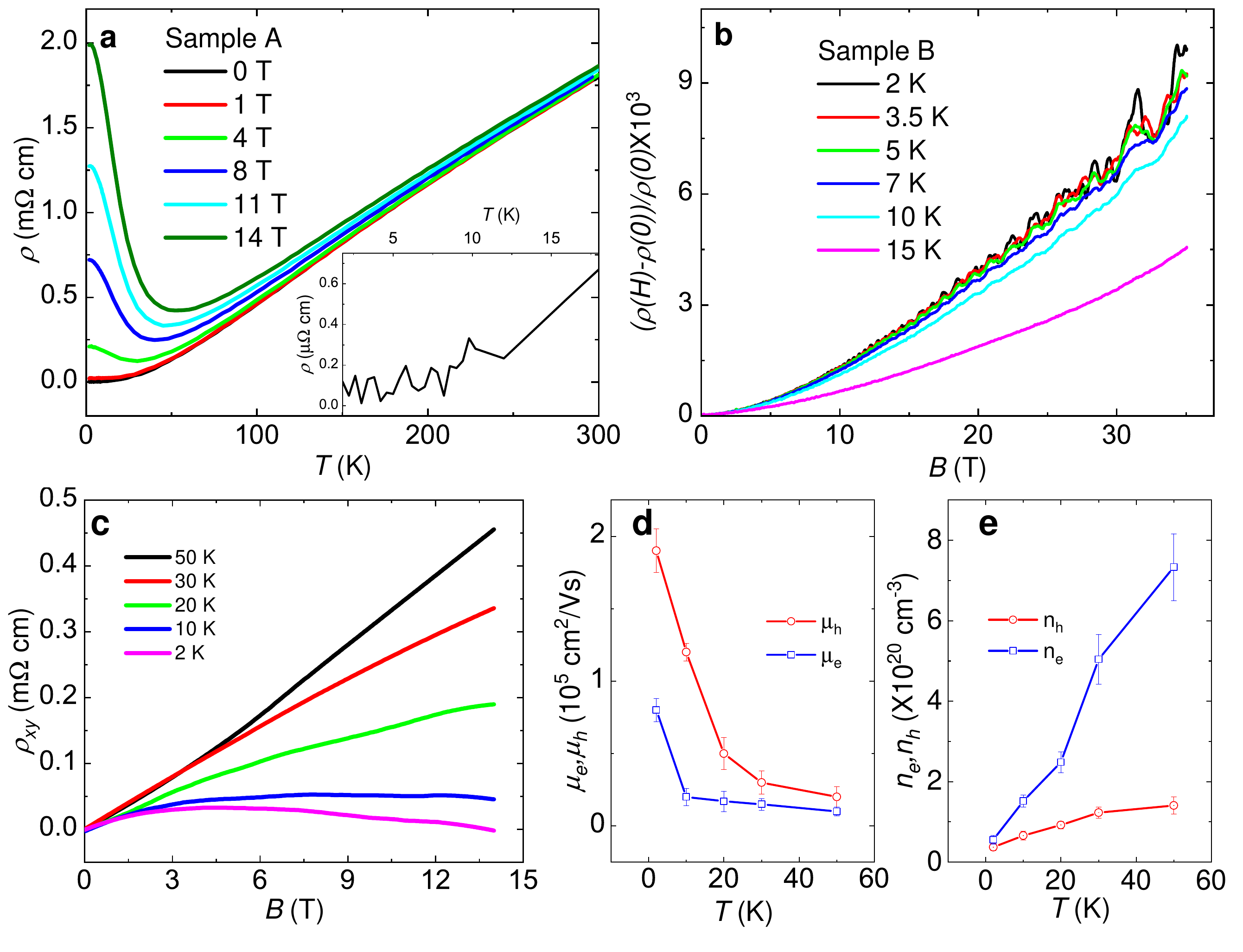}
		\caption{\textbf{Extremely large magnetoresistance and carriers of Dirac semimetal BaAl$_4$.} 
			\textbf{a}, Temperature dependence of the resistivity at different magnetic field which is perpendicular to the electrical current for single crystal sample A.
			\textbf{b}, Transverse magnetoresistance with current in the $ab$ plane and field along the $c$ axis, for single crystal Sample B at several representative temperatures under field up to 35 T. In the measurements, the current is in the $ab$ plane and the field is along the $c$ axis. 
			\textbf{c}, Field dependence of the Hall resistivity measured at several representative temperatures. \textbf{d--e,} Temperature dependence of the mobility of the holes ($\mu_h$) and electrons ($\mu_e$), as well as the density of the holes ($n_h$) and the density of the electrons ($n_e$). }
	\end{center}
\end{figure*}

\begin{figure*}[tbp]
	\includegraphics[scale=0.5]{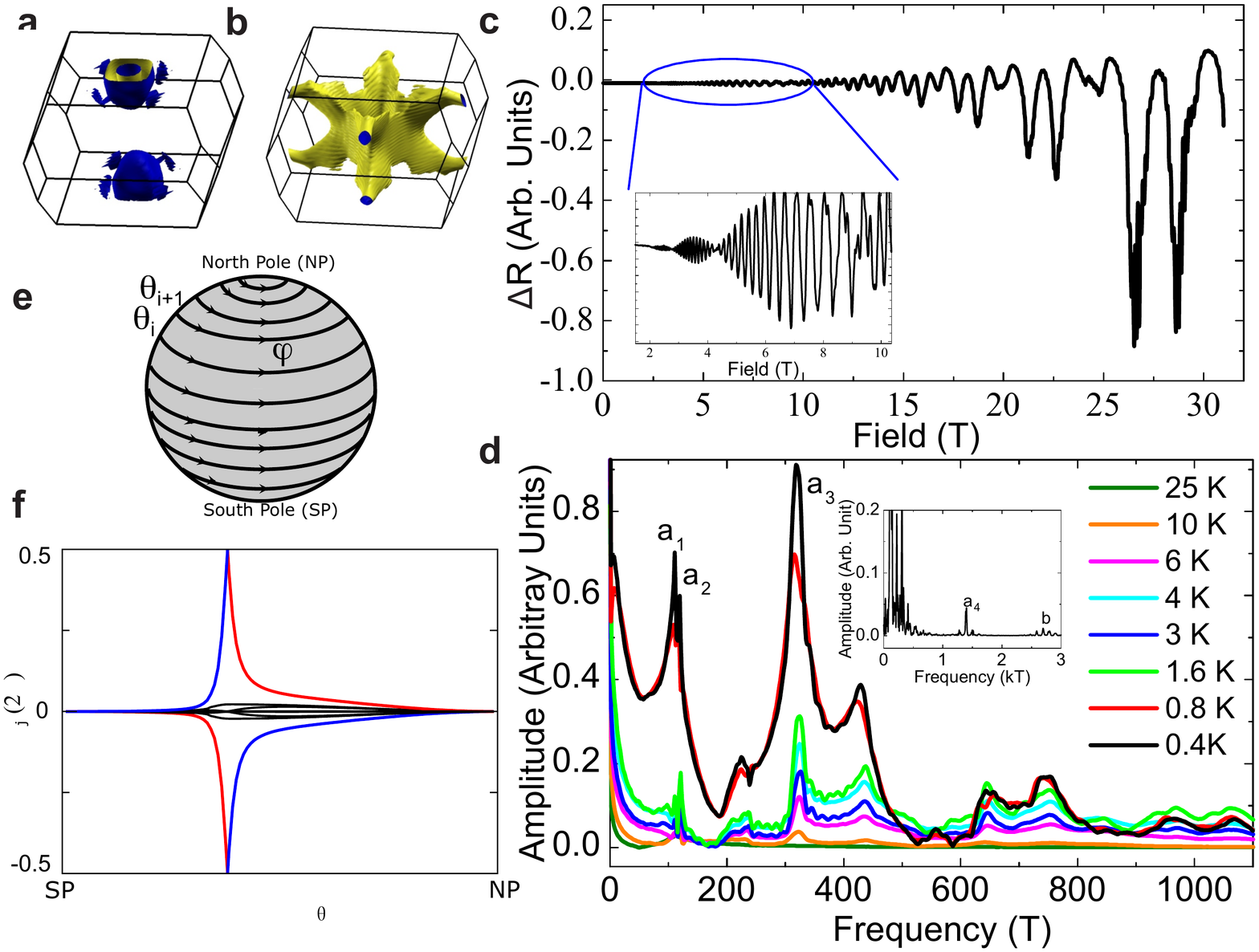}
	\caption{\textbf{Fermi surface and quantum oscillation of Dirac semimetal BaAl$_4$.}
		\textbf{a-b,} Calculated hole pockets (a) and electron pockets (b) in the Fermi surface for BaAl$_4$ by density function theory with spin-orbit coupling.
		\textbf{c,} The quantum oscillation in the magnetoresistance of BaAl$_4$ single crystal at 2K. The inset shows the FFT spectrum of the quantum oscillation.
		\textbf{d,} Enlarged part of the low frequency region of the FFT spectrum of the quantum oscillation for Sample B of BaAl$_4$ at different temperatures. The inset shows the detail of the FFT spectrum in high frequency region at 2K. (e) Wilson loop spectrum on a spherical surface enclosing the Dirac point $(0, 0, k_z^c)$. (f) Phase $\varphi_j$ of the individual loop eigenvalues $(e^{i\varphi_j})$ as a function of the azimuthal angle $\theta$.}
\end{figure*}

\subsection{Magnetoresistance and Quantum Oscillations}

Fig. 4 presents the electrical transport properties of single-crystalline BaAl$_4$ samples. Extremely high mobilities approaching 200,000~cm$^2$/Vs (see Hall analysis below) are reflective of the high quality of the samples, also indicated by very large residual resistivity ratios (RRR=$\rho(300K)/\rho(2K)$=1,500 in Sample A) and tiny residual resistivities ($\sim 90$ n$\Omega\cdot$cm).
As observed in other TSMs such as WTe$_2$, NaBi$_3$, Cd$_3$As$_2$, TaAs, LaBi and other related compounds~\cite{Cd3As2-1,WTe2-MR,Liang_2014,TaAs-Chen,LaSb_Tafti}, a dramatic onset of strong MR develops at low temperatures and presents astonishingly high and unsaturating MR values attributed to the opening of an energy gap by magnetic fields at the band-touching point.  In BaAl$_4$, we also observe similar behavior, with a large upturn in resistivity at low temperatures that tends to saturate.
Fig. 4(b) presents the transverse MR for Sample B with current applied parallel to the $ab$ plane and field along the $c$ axis, showing a quadratic evolution with field that does not saturate up to at least 35~T. MR exhibits an amplitude correlated with crystal quality (shown in Fig. S2 of supplementary materials). For Sample A with the highest $RRR\sim1500$, the MR approaches $2\times10^5\%$ at 14~T (shown in Fig. 4(a) and also in Fig. S2 of supplementary materials), a value comparable to that reported in Cd$_3$As$_2$, the TaAs family and other TSMs (including LaP with P=Sb and Bi, WTe$_2$, NbSb$_2$ family). At higher fields, the MR does not saturate, as shown for Sample B data up to 35~T, where MR approaches $10^6\%$ (FIG. 4(b)). 



As shown in Fig. 4(c), the Hall resistivity $\rho_{xy}$ is positive above 50 K and nearly linear with field, indicating holes as the majority carriers at high temperatures. However, $\rho_{xy}$ deviates from linear at temperatures below 50 K, and even changes sign at low temperature and high field, indicating a coexistence of electrons and holes. It has been proposed that the extremely large MR in several semimetals arises due to a perfect compensation between electron and hole carrier concentrations~\cite{WTe2-Weyl}. The calculated band structure of BaAl$_4$ does indeed show complex coexistence of electron and hole pockets. The hole pockets (see Fig. 5(a)) corresponding to the Dirac point along $\Gamma-Z$ direction mostly consist of two separated hollow cylinders surrounding the $Z$ points, while the electron pocket is a one-piece structure centered at the $\Gamma$ point shown in Fig. 5(b). However, the volume surrounded by the hole and electron pockets are different. To estimate the carrier densities and mobility, a semi-classical two-band model is used to analyze the longitudinal conductivity $\sigma_{xx}(B)$ and Hall conductivity $\sigma_{xy}(B)$ in the magnetic fields $B$:
\begin{eqnarray}
\sigma_{xx}(B)&=\frac{n_eq\mu_e}{1+(\mu_eB)^2}+\frac{\sigma_{xx}(0)-n_eq\mu_e}{1+(\mu_hB)^2},\\
\sigma_{xy}(B)&=\left[\frac{n_e\mu_e^2}{1+(\mu_eB)^2}-\frac{n_h\mu_h^2}{1+(\mu_hB)^2}\right]eB.
\end{eqnarray}
Here $n_e$ ($\mu_e$) and $n_h$ ($\mu_h$) are the carrier densities (mobilities) of electrons and holes, respectively. $\sigma_{xx}(0)$ is the longitudinal conductivity at 0 T. The typical fitting results at several different temperatures are shown in Fig. S3 (a) and (b) in the Supplementary Materials, and the fitting parameters are shown in Fig. 4(d,e) in temperature range 2 K $\sim$ 50 K. At low temperatures, holes are the minority carriers, but their mobility is nearly one order higher than the mobility of electrons. At 2 K, $\mu_h$ approaches $\sim$ 20 m$^2$/Vs. With increasing temperature, $\mu_h$ decreases significantly and becomes comparable to $\mu_e$. Hence, the minority hole carriers dominate the transport in the entire temperature range because of their high mobility. This is consistent with the calculated electronic structure, where the Dirac pockets have hole characteristics.

Quantum oscillations measurements further clarify the nature of carriers in BaAl$_4$. As shown in Fig. 4(b), the measured MR of several BaAl$_4$ crystals exhibit clear Shubnikov-de Haas oscillations. Subtracting a background MR yields an oscillatory signal composed of several components observable down to a few Teslas, as shown in Fig. 5(c). The oscillation frequencies of different pockets in the calculated Fermi surface are extracted by finding the supercell k-space extremal area implemented in the SKEAF code~\cite{skeaf} and are compared to the frequencies by Fourier transform analysis of the experimental curves, correspondingly the oscillation frequencies got by Fourier transform of our experiment curves (in FIG. 5(d)) are assigned to different pockets the calculated Fermi surface.  Our experimental results are in consistent with the calculated Fermi surface shown in Fig. 5(a-b), and also most of them are close to the previous report reports~\cite{SrAl4_FS}, as shown in Table IV of SM. In the high frequency region as shown in the inset of Fig. 5(d), the frequency $F_{\beta}\sim2.5$ kT corresponds to the electron pocket shown in Fig. 5(b), and the other frequency $F_{\alpha_4}\sim1.6$ kT corresponds to the outer branch of the hole pocket surrounding $\Gamma$ point in Fig. 5(a). A small frequency $F_{\alpha_3}\sim300$ T corresponds to the inner branch of the hole pocket.  Besides these frequencies already observed in Ref. \cite{SrAl4_FS} where a detailed angular dependence of these frequencies consistent with calculated Fermi surface is reported, two very close oscillation frequencies $F_{\alpha_1}\sim100$ T and $F_{\alpha_2}\sim106$ T which are missing in previous report\cite{SrAl4_FS}, are observed. These two close frequencies give a clear beating behavior shown in Fig. 5, and correspond to the small hole pockets along $\Gamma-X$ lines. Lifshitz-Kosevich analysis of the temperature-dependent oscillation amplitude (Fig. S4 of SM) for these two frequencies $F_{\alpha_1}$ and $F_{\alpha_2}$ reveals a similar effective mass of $\sim0.2(1)m_e$. This small effective mass of the hole pockets is consistent with the calculated hole-like Dirac spectrum (with a mass of about $0.15m_e$, shown in Table IV of SM) and the ARPES measurement.


\subsection{Discussion and Conclusions}

Quantum oscillations, Hall resistivity and ARPES results reveal that although the electron and hole pockets coexist in BaAl$_4$, they harbor different densities. This implies that some other mechanism beyond the perfect compensation between electron and hole pockets should govern the extremely large MR found in BaAl$_4$, which may be related to the Dirac spectra as discussed for other TSMs. The extremely small residual resistivity $\sim 90$ n$\Omega$ cm is quite anomalous given the low carrier density $\sim 10^{20}$ cm$^{-3}$, and is similar to the case for other topological semimetals such as WTe$_2$ ($\sim 30 n\Omega\cdot cm$) and La(Bi,Sb) ($\sim 0.1\mu\Omega\cdot cm$), suggesting a topology-derived mechanism for reduction in scattering and large MR~\cite{Liang_2014,WTe2-SpinTexture}. Although the Fermi surface is about $0.4$ eV away form the Dirac points, the winding feature of the Wilson loop spectrum can still b stabilized by the $C_{4z}$ symmetry.  FIG. 5 shows the Wilson loop spectrum on a spherical surface enclosing the Dirac point $(0, 0, k_z^c)$ (e) and the phase $\varphi_j$ of the individual loop eigenvalues $(e^{i\varphi_j})$ as a function of the azimuthal angle $\theta$ (f).  On a given-$\theta$ Wilson loop, $(r\dot sin(\theta)cos(\phi), r\dot sin(\theta)sin(\phi), k_z(\theta))$ with $k_z(\theta)=k_z^c+r\dot cos(\the\theta)$ and the spherical radius $r$, the Wilson loop is mapped as $C_{4z}W_{2\pi\leftarrow0}C_{4z}^{-1}=W_{\frac{5\pi}{2}\leftarrow\frac{\pi}{2}}=W_{\frac{\pi}{2}\leftarrow0}W_{2\pi\leftarrow0}W_{0\leftarrow\frac{\pi}{2}}$, where $W_{B\leftarrow A}$ represents a parallel transport from $\phi=A$ to $\phi=B$. Then we can redefine a symmetry $\varsigma_{4z}=W^{-1}_{\frac{\pi}{2}\leftarrow0}C_{4z}$, which yield $\varsigma_{4z}W_{2\pi\leftarrow0}\varsigma_{4z}^{-1}=W_{2\pi\leftarrow0}$. Thus, each Wilson energy $\varphi_j$ can be labeled by $\lambda_{\alpha}e^{-i\frac{\varphi_j}{4}}$ with $\lambda_j=e^{i\frac{(2n-1)\pi}{4}}$ for $j=\{1,2,3,4\}$, where we use $\varsigma_{4z}=W^{-1}_{\frac{\pi}{2}\leftarrow0}C_{4z}=-e^{-i\varphi_j}$. Our further calculations show that the $\varsigma_{4z}$ eigenvalue of the blue(red) line is $\lambda_2e^{-i\frac{\varphi_j}{4}}$ ($\lambda_3e^{-i\frac{\varphi_j}{4}}$). Thus, those two Wilson eigenstates can not hybridize with each other, giving rise to a stable crossing in the Wilson loop spectrum. As long as the surface states are $C_{4z}$ symmetric, the winding feature of the Wilson loop spectrum is stable. This could, for instance affect the bulk transport properties, but requires further investigation.

Our results indicate BaAl$_4$ as a new topological semimetal, as confirmed by calculations, symmetry analysis, photoemission and transport experiments.
Since a large diversity of physical phenomena appears in compounds derived from this prototype structure, one can consider the MA$_4$ family as a new playground for exploring the interaction between the non-trivial topology and symmetry-breaking or correlated phases, including superconductivity, heavy fermion and quantum critical systems, charge/spin density wave, etc.. It could also be expected that some distortion/modification of the original crystal structure in the derived structures will slightly modify the topological electronic structure and induce new phenomena. One interesting direction is to consider the consequences of breaking the inversion symmetry to achieve a non-centrosymmetric structure (e.g. LaPtSi$_3$ and other materials), where band splitting could be expected to achieve Weyl or multiple-fermion states~\cite{Armitage2018,}. Noncollinear magnetic order (such as in EuAl$_4$) may also induce band splitting, yielding other exotic topological states. Extending the symmetry analysis to the rest of the MA$_4$ family of materials (M=Sr, Ba, Eu; A=Al,Ga,In), as well as the enormous set of derivative structures\cite{TMDatabase}, will expose new topological materials.


\section{Methods}
\subsection{Crystal growth}
Single crystals of BaAl$_4$ were synthesized by a high-temperature self-flux method. Chunks of Ba ($99.98\%$, Alfa Aesar) and pieces of Al ($99.999\%$ metal basis, Alfa Aesar) in the
ratio of 2:98 were placed in the Canfield alumina crucible set with a decanting frit (LSP ceramics), sealed in the quartz ampule under partial Ar pressure, heated to 1150 $^o$C, held at that temperature for 6 h, cooled down at 3 degree/h to 950 $^o$C with subsequent decanting of the excess Al with the help of a centrifuge. The crystals tend to grow with large faceted surfaces normal to the (001) crystallographic plane. X-ray diffraction data were taken at room temperature with Cu K$_{\alpha}$ ($\lambda=0.15418$ nm) radiation in a powder diffractometer.

\subsection{Transport measurement}
Electrical transport measurements up to 14 T were conducted on polished samples in a Quantum Design Physical Properties Measurement System (Dynacool) applying the conventional four-wire contact method. High field MR and dHvA oscillation were measured at the National High Magnetic Field Laboratory (NHMFL) Tallahassee DC Field Facility up to 35 T, using a one-axis rotator in a He-3 cryostat with temperature range 0.35 K $\sim$ 70 K. In all transport measurements, the electrical current flowing in the $ab$-plane. The magnetization and dHvA oscillations were measured using a piezoresistive cantilever. The magnetic field  rotated from a direction parallel to the $c$-axis to a direction parallel to the $ab$-plane.

\subsection{Photoemission measurement}
ARPES measurements on single crystalline samples of BaAl$_4$ were performed at the Beamline 4.0.3 (MERLIN) of the Advanced Light Source in Berkley, California. 
Samples were cleaved in situ to yield clean (001) surfaces and measured at 14--20 K in ultra high vacuum better than  $3\times10^{-11}$~Torr using photon energy of 80-120~eV with Scienta R4000 analyzer. The energy resolution was 20-30~meV and the angular resolution was better than $0.2^{\circ}$ for all measurements.
The inner potential $V_0=10.5$ eV was used to calculate the momentum perpendicular to the surface.

\subsection{DFT calculation}
Theoretical calculations compared to ARPES were performed within generalized gradient approximation (GGA) \cite{gga} for bulk BaAl$_4$ as implemented in the VASP package \cite{vasp-1,vasp-2,vasp-3}, using a relativistic version of PAW potentials\cite{paw-1,paw-2}. The experimental crystal data ($a=b=4.566$\AA and $c=11.278$\AA) were used for the bulk calculation. PAW pseudopotentials with a plane wave energy cut-off of 40 Ry were used, and Monkhorst k-mesh was chosen as $24\times24\times24$. All calculations prepared by VASP were double checked by the QUANTUM ESPRESSO Package\cite{QE-2009,QE-2017}. 

\section{Author contributions}
J.P., K.W., and A.L. initiated and directed this research project.
The samples were grown and characterized by K.W.
Transport measurements were carried out by K.W. with the assistance of D.E.G.
R.M., J.H.M, and D.W.L. carried out ARPES measurements with the assistance of J.D., and R.M. analyzed the data.
L. W and R. M did a portion of the calculation. Z.W. and B.A.B did the calculation and symmetry analysis.
K.W. and R.M. wrote the text, with feedback from all authors.
All authors contributed to the scientific planning and discussions.

\section{acknowledgments}
This work was supported by the Gordon and Betty Moore Foundation's EPiQS Initiative through Grant GBMF4419.
ARPES work and a portion of the calculation are supported by the U.S. Department of Energy, Office of Science, Office of Basic Energy Sciences, Materials Sciences and Engineering Division under Contract No. DE-AC02-05-CH11231 (Quantum materials KC2202).
A.L. acknowledges partial support for this research from the Gordon and Betty Moore Foundation's EPiQS Initiative through grant GBMF4859.
A portion of this work was performed at the National High Magnetic Field Laboratory, which is supported by National Science Foundation Cooperative Agreement No. DMR-1157490 and the State of Florida.
R.M. also acknowledges support from the Funai Foundation for Information Technology. Z.W. and B.B. are supported by the Department of Energy Grant No. de-sc0016239, the National
Science Foundation EAGER Grant No. noaawd1004957,
Simons Investigator Grants No. ONRN00014-14-1-0330,
and No. NSF-MRSEC DMR- 1420541, the Packard
Foundation, the Schmidt Fund for Innovative Research. Z.W. acknowledges support from the CAS Pioneer Hundred Talents Program.
\bibliography{./TopologicalMaterials}
\bibliographystyle{naturemag}

\newpage

\end{document}